# The 1908 Tunguska event in data of the nearest weather station

Andrei Ol'khovatov
https://orcid.org/0000-0002-6043-9205

Independent researcher
(Retired physicist)

Russia, Moscow
email: olkhov@mail.ru

**Dedicated to the blessed memory of my grandmother ( Tuzlukova Anna Ivanovna ) and my mother ( Ol'khovatova Olga Leonidovna )**

**Abstract.** This paper is a continuation of a series of works, devoted to various aspects of the 1908 Tunguska event. In this paper data is discussed from a weather station in the settlement of Kezma. This weather station was the nearest one to the epicenter of the 1908 Tunguska event. The station was about 215 km southwards from the epicenter. Also eyewitness accounts from Kezhma and some places on the Angara River are considered.

The paper demonstrates the diversity of optical manifestations of the 1908 Tunguska event reported from Kezhma and its surroundings. A note by the observer of the weather station in Kezhma, concerning, in particular, the appearance of the two huge fiery circles is very important.

The weather log of the weather station shows that the Tunguska event occurred during the sharp increase in cloud cover and deteriorating weather. At the beginning of July, rains and thunderstorms, including hail, began according to the weather log.

In the opinion of the author, the data of the Kezhma weather station and the eyewitness accounts are in agreement with the geophysical interpretation of the 1908 Tunguska event.

In any case, the meteorological factor is an important aspect of the Tunguska event, and so it deserves detailed research.

## 1. Introduction

This paper is a continuation of a series of works in English, devoted to various aspects of the 1908 Tunguska event [Ol'khovatov, 2003; 2020a; 2020b; 2021; 2022; 2023a; 2023b; 2025a; 2025b; 2025c; 2025d; 2025e; 2025f; 2025g; 2025h; 2025i; 2026a; 2026b]. The works can help researchers to verify the consistency of the various Tunguska interpretations with actual data. A large number of hypotheses about its causes have already been put forward. However, so far none of them has received convincing evidence. As it is written on the title web-page of the web-site created by KSE (see below about KSE) tunguska.tsc.ru/ru/ (translated by A.O.):

> "About a thousand researchers have devoted years of their lives to the Tunguska phenomenon. However, there is still no well-founded scientific understanding of what happened over the Siberian taiga on June 30, 1908."

This is probably why new hypotheses appear almost every year, not only in the mass-media, but also in scientific literature. At the same time, any hypothesis should not contradict the known facts about the event. Unfortunately, the authors of new hypotheses, as well as the authors of popular science articles, often use data, many of which turned out to be not entirely accurate, or even incorrect (some examples can be read in [Ol'khovatov, 2025c]). The author of these papers hopes that the papers will help both the authors of various hypotheses and their readers to evaluate the validity of the proposed hypotheses.

Let's start with brief info about research of the Tunguska event. The Committee on Meteorites of the USSR Academy of Sciences (KMET) stopped research the area of the Tunguska event in the early 1960s. Later amateurs (consisting mainly of scientists, engineers and students) most of whom united under the name Kompleksnaya Samodeyatel'naya Ekspeditsiya (KSE) continued research (KSE started research in 1959). Since the late 1980s foreign scientists take part too.

Please pay attention that so called the epicenter of the Tunguska event is assigned to 60°53' N, 101°54' E.

In this paper its author (i.e. A.O.) for brevity will be named as “the Author”.

## 2. The weather station data

The weather station closest to the epicenter of the Tunguska event was in the settlement of Kezhma (Kezhemskoe) at 58°58' N, 101°04' E [Rykatchew, 1911]. In other words, the station was about 215 km to the south from the epicenter. The location is shown in Fig. 1. In Fig.1, the red arrow shows the trajectory of the alleged Tunguska spacebody, based on eyewitness accounts (obtained in the 1960s) and the position of the axis of symmetry of the forestfall - see details, for example, in [Ol'khovatov, 2023a].

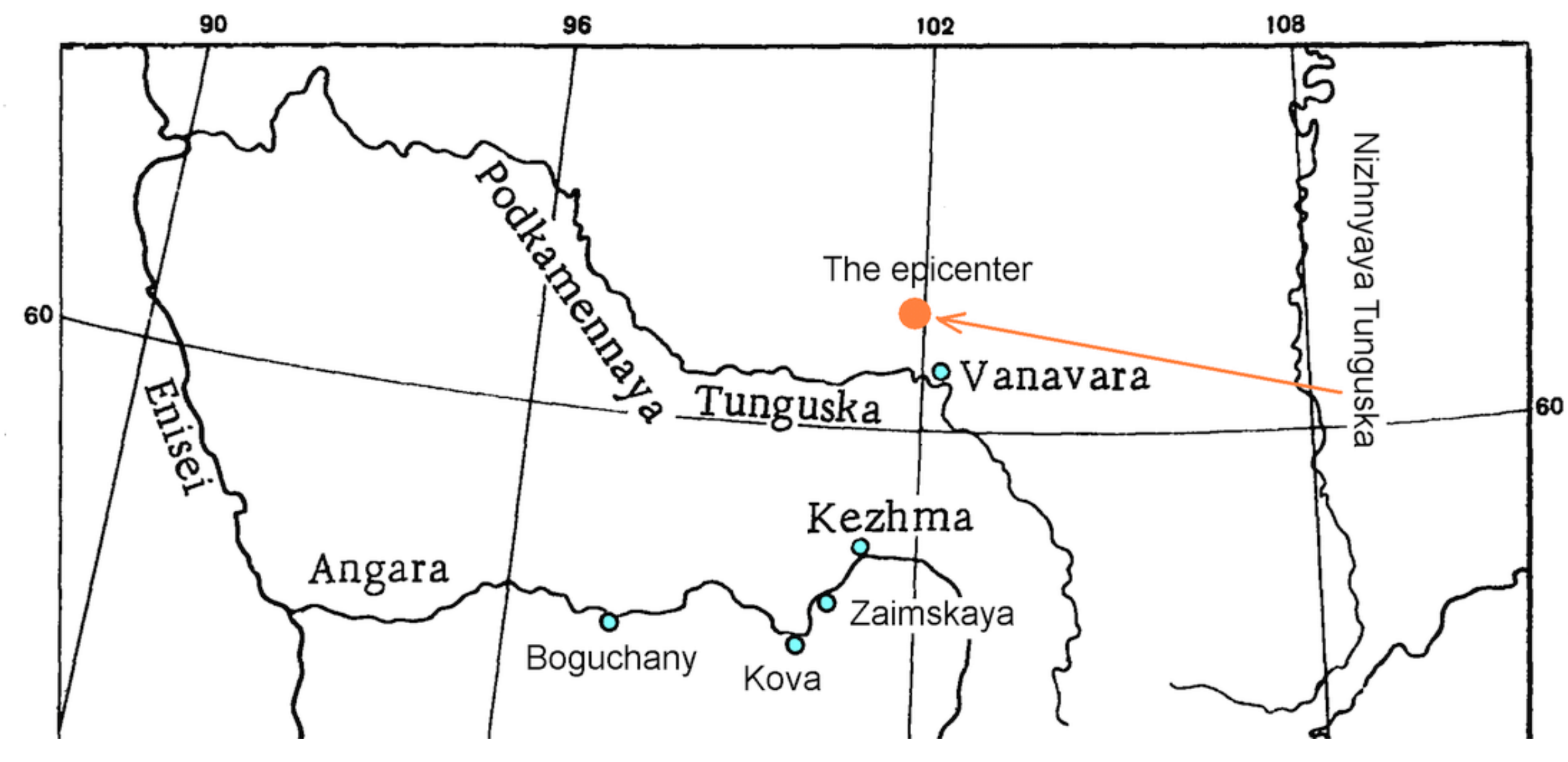

**Fig.1**

The Kezhma weather station was a second-category station and did not have a barometer unfortunately. Measurements at the station were carried out 3 times a day - at 7 a.m., at 1 p.m. and at 9 p.m. local time.

The observer of the station was Anfinogen Kesarevich Kokorin. He was interested in the Tunguska event, collected info, etc. In the late 1920s he even helped to L.A. Kulik. More info about A.K. Kokorin can be read in [Ol'khovatov, 2023b], for example.

In 1933 I.S. Astapovich published some minor weather data of the Kezhma station (see translation in English in [Astapowitsch, 1940]). In there he also published a very important note by A.K. Kokorin regarding the Tunguska event. Here it is from [Astapowitsch, 1940]:

> "At Kezhma on the Angara (there is still another, Upper Kezhma, also on the Angara, at Bratsk), A. K. Korin adds to the remarks on the

> meteorological observations for June 30, 1908 (archives of the former Irkutsk magnetic and meteorological observatory): “At 7 a.m., there appeared in the north two fiery circles of colossal dimensions; $4^m$ after their appearance, the circles vanished; soon after the disappearance of the fiery circles, was heard a powerful noise, resembling the sound of wind, which went from north to south; the noise lasted about $5^m$. Then followed sounds and crashes resembling the discharges of huge cannon, which made the windows rattle. These shots continued for $2^m$ and then was heard a crackling, resembling rifle shots. These latter continued for $2^m$. All this happened when there was a clear sky.””

Unfortunately there is a mistype in the A.K. Kokorin's surname. As his account is rather important, here is another translation (by A.O.):

> “June 30…new calendar at 7 o'clock in the morning, two huge fiery circles appeared in the north; after 4 minutes from the beginning of the appearance, the circles disappeared; soon after the disappearance of the fiery circles, a strong noise was heard, similar to the noise of the wind, which went from north to south; the noise lasted about 5 minutes. Then the sounds and crackling followed, similar to the shots from huge guns, from which the frames trembled. These shots lasted for 2 minutes, and after them there was a crackling sound, similar to a shot from a gun. These last lasted 2 min. Everything that happened was under a clear {see below about the translation - A.O.} sky”.

This account is remarkable in many aspects (absence of any super-bolide and even its trail, etc.). Also this account in rather reliable (at least comparing with many other accounts), as it was written by the 'official' trained observer in official document right after the Tunguska event. Unfortunately, this account is rarely found in works devoted to the Tunguska event.

It is important to note that Kokorin did not use the word "bezoblachnoe", i.e. "cloudless" (sky), but used instead the word "yasnom" (sky), which can mean not only "cloudless", but also "light"/"sunny" (sky). A possible reason of this is discussed below.

Figure 2 shows a graph of cloudiness from June 27 to July 3, 1908 according to the station data. Data for 7 a.m. on June 30 is highlighted in larger size and red.

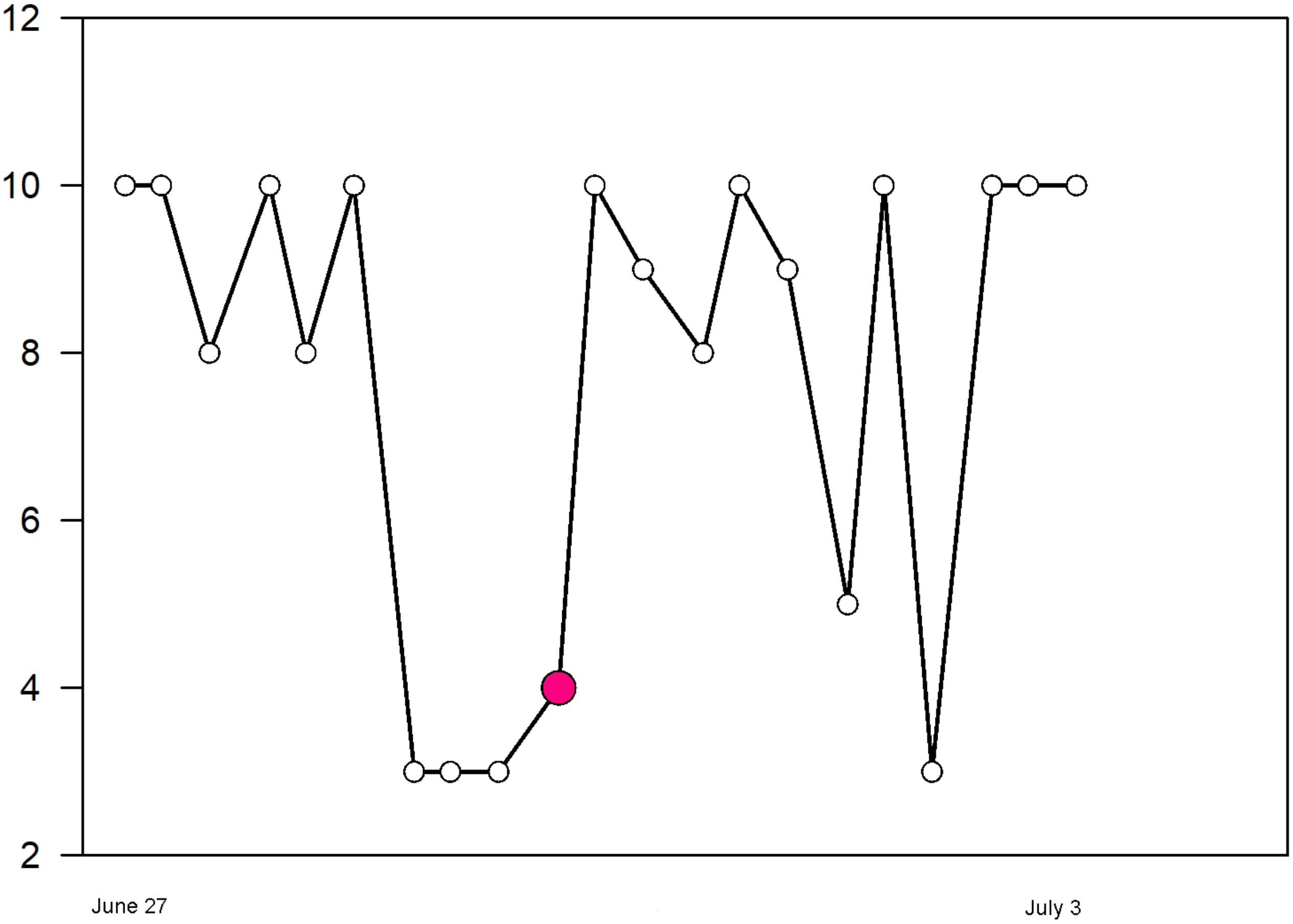

**Fig.2**

As it can be seen from Fig.2, cloud cover began to increase rapidly around 7 a.m. (i.e. about the time of the Tunguska event). Perhaps this is why Kokorin chose not to use the word "cloudless".

In any case, let's take a closer look at this sharp increase in cloudiness. There is an account that the sky near Kezhma was cloudless for at least several hours before the Tunguska explosion. Here is a fragment from the account [Vasil'ev, et al., 1981], translated by A.O.:

> "A member of the management of the Consumers Association of Kezhma, I. K. Vologzhin, reported the following to L.A. Kulik on the 21st of November, 1921: "In the year 1908, on June 17, I was about 20 versts above the village Kezhma (along the Angara river) in the place Chirida. In the morning we examined our {probably fishing - A.O.} nets. It was a clear morning. There was not a cloud. With me was an old man. We heard - a thunder, then - another one, and gradually the thunder rumblings began to diminish toward the north. I don't remember what time it was, but the sun hadn't risen yet, it was just dawn. In the winter the inhabitants of the village

Kezhma, who traded around the Podkamennaya Tunguska...””

By the way, it follows from the account that something peculiar took place in the region several hours before the Tunguska explosion.

Now let's consider time near the Tunguska explosion. The following newspaper article is often quoted in publications on the Tunguska event. Here is a fragment from the article published in the newspaper “Krasnoyarets” in the issue of July 13, 1908 (the Julian calendar) - see more info in [Ol’khovatov, 2023a]:

> “S. {settlement - A.O.} Kezhemskoe. On June 17th (this month), an extraordinary atmospheric phenomenon was noticed in this region.
>
> <...> The sky, at first glance, was completely clear. There was no wind or clouds. But with close observation, in the north, i.e., where the blows seemed to be heard - on the horizon, something like an ash-colored cloud was clearly noticed, which gradually decreasing, it became more transparent, and by 2-3 o'clock in the afternoon it completely disappeared. <...>
>
> According to eyewitnesses, before the first blows began, some kind of heavenly body of a fiery kind cut through the sky from the south to the north with a tendency to NE, but for the speed (and most importantly - the surprise) of the flight, neither the size nor the shape of it could be seen. But on the other hand, many and in various villages perfectly saw that with the touch of the flying object to the horizon, in the place where abovementioned peculiar cloud was subsequently noticed later, but much lower than the location of the latter - at the level of forest tops, it was as if a huge flame broke out, splitting the sky."

It is noteworthy that the flying object left no trail behind.

Basing on such observations a police officer reported that "On the 17th of last June, at 7 o'clock in the morning over the village of Kezhemskoye (on the Angara) from the south towards the north, in clear weather, a huge aerolite flew high in the sky, ..." - see [Ol’khovatov, 2023a].

According to the newspaper story, there was something like an ash-colored cloud on the horizon about the time of the Tunguska event.

By the way, the station registered at 7 a.m. wind 3 m/s. Possible reasons for the discrepancy lay not only in the different locations of the observation site (see below), and perhaps in the difference in the time of observation, but also in the fact that the weather vane of the meteorological station was at a height of ~8.5 meters.

It was written in the article of the newspaper “Krasnoyarets” that "many and in various villages perfectly saw that with the touch of the flying object to the horizon..."- let's look attentively what was seen in another village not far from Kezhma. Here is from [Vasil'ev, et al., 1981], translated by A.O.:

"I.A. Kokorin, a resident of the village of Kezhma, interviewed by E.L. Krinov in 1930.

"Together with Bryukhanov and others (5-6 people), I was traveling by boat along the Angara River to the village of Kova to collect millstones. Near the village of Zaimskaya, we approached the shore and, having anchored the boat at the shore, went "up the hill" to the village located directly to the south. Having walked a few steps away from the boat, we saw to our right (directly to the west) a fiery red flame flying at an angle to the ground to the north. It spread out, as if fired from a gun, three times larger than the sun, but no brighter: it was possible to look at it, and we saw how the flame disappeared behind the mountains to the northwest. We noticed the flame when it had already appeared in the sky. As soon as the flame touched the ground, sounds like continuous cannon fire were heard. The sounds lasted for no more than half an hour. The ground shook as the sounds were heard, and the windows rattled and continued to rattle as we entered the house. The water in the river was calm. <...>".

Bryukhanov T.I., mentioned in the previous message, was interviewed by E.L. Krinov in 1930 in the village of Kezhma.

“We were lying in the boat and, about 200 sazhen'{i.e. about 140 meters - A.O.} from the shore, we saw in front of us in the northwest rays kosikom {meaning is not clear, possibly similar to a triangle - A.O.}, with the wide end down, flying to the north. Having reached the ground, they disappeared behind the forest, and in their place, in the strip of sky along which the rays had flown, many separate puffs of smoke formed. When the rays disappeared, small waves formed on the water. After that, we landed on the shore, tied up the boat, and headed for the village of Zaimskaya. We had just entered the house and said hello when we heard loud sounds, like gunshots, which didn't last long. The ground and the windows were shaking at this moment. The rays were fiery red, bright, but it was possible to look at them without pain in the eyes. I didn't notice how or when the smoke disappeared. The rays appeared (as Bryukhanov indicated with his hand) at an altitude of about 60 degrees"".

The village of Zaimskaya is located approximately 40 km in a straight line on azimuth ~219° (i.e. to the south-west) from Kezhma.

The accounts indicate that the flame/rays has a very specific shape and moderate brightness. It appears that its angular velocity across the sky was also not very high.

What is even more remarkable is that its trajectory passes to the west of Zaimskaya. By the way, according to I.A. Kokorin, the flame disappeared behind the mountains to the northwest.

It follows from this that the object observed from Zaimskaya and the object observed from Kezhma (published in the newspaper "Krasnoyarets") are different objects.

By the way, could "many separate puffs of smoke formed" be the initial stage of cloud formation?

Anyway, let's consider what was seen near-by. Here is from [Vasil'ev, et al., 1981], translated by A.O.:

> "Bryukhanov D.F., interviewed by L.A. Kulik in 1938, said:
> "At that time I was plowing arable land on Narodimaya {spelling? - possibly Narodimyi - A.O.} (6 km. to the west of Kezhma), when I sat down to breakfast near my plow, suddenly there were blows, like cannon shots. The horse fell to its knees. Flame flew out from the north side over the forest. I thought: the enemy is shooting (at that time they were talking about the war). Then I see the spruce forest bent down: I think about the hurricane, grabbed the plow with both hands so that it wouldn't carry. The wind was so strong that it blew a little soil off the surface of the earth; and then this hurricane drove the water on Angara with a {big -A.O.} wave: I could see everything well, because the arable land was on a hillock.""

Remarkably that the eyewitness saw just the flame. Neither the alleged Tunguska spacebody superbolide, nor even its trail, nor "many separate puffs of smoke formed".The latter means that the "many separate puffs of smoke formed" seen from Zaimskaya (as well as the flame/rays) were at relatively low altitude. By the way, the strong windstorm is also remarkable.

By the way, regarding the windstorm. Here the Author just wants to point to some analysis by other researchers. Statistical analysis of about 700 accounts (collected in various years) conducted in [Demin et al., 1984] revealed that (translated by A.O.):

> "The Tunguska phenomenon, according to eyewitnesses, was accompanied by various meteorological phenomena (Table 7). The most frequently observed were "strong wind", as well as "haze, fog, fog". The complexity of atmospheric processes is evidenced by sharp changes in air temperature recorded in a number of accounts. Thunderstorms, individual lightning discharges, local development of windstorms, hurricanes and whirlwinds are noted."

Now let's find out what was seen west of Zaimskaya. Here is a fragment from an account from [Vasil'ev, et al., 1981], translated by A.O.:

> "S.I. Privalikhin, 39, interviewed by E.L. Krinov in 1930, resident of the village of Kova (on the Angara River).

"... on a perfectly clear day, in the morning after tea. The sun had already risen quite high. I was about 15 years old at the time. I was 10 verst {~10 km - A.O.} from the village of Kova, in a ploughed field. I had just managed to harness the horses to a harrow and was standing to tie up another one when suddenly I heard a faint shot from a gun (one blow) to my right. I immediately turned and saw something flying, as if on fire, elongated: its forehead broader, its tail narrower, the color as of fire in daylight (white), many times larger than the sun, but much dimmer, so that it could be seen. Behind the flame, a kind of dust remained: it swirled in clumps, and blue streaks remained from the flame. It flew quickly, for about three minutes. The flame disappeared behind the mountain range between the north and west (slightly west of north). I saw it flying a little below the distance between the zenith and the horizon, above the summer sunset. As soon as the flame disappeared, sounds louder than gunshots were heard. Tremblings of the ground were felt, and glass rattled in the windows of the winter hut, where I ran immediately after seeing the flame.""

The village of Kova is at ~58° 18' N, ~100° 20' E. It is noteworthy that in both villages the objects (or the object?) were visible on the western side. The description of the flying object seen near Kova is a bit reminiscent of the one in Zaimskaya. So it can't be ruled out that it was the same object.

However farther west from Kezhma near the village of Boguchany (~58°23′ N, ~97°26′ E) a reported object was different. Here is a fragment from an account from [Vasil'ev, et al., 1981], translated by A.O.:

"Information reported to L.A. Kulik on January 19, 1924 by engineer V.P. Gundobin, who lived for two years on Podkamennaya Tunguska near the area where the meteorite fell [3]: <...>

"The most interesting fact is the report of Ivan Vasilyevich Kokorin. He was sailing along the Angara River at the time (traveling with boats). This was at the Murskii Rapid (near the village of Boguchany - authors) at 5:00 a.m. on June 17, 1908. He was at the helm of the boat. His impressions were as follows: a bluish light flashed in the north, and a fiery body, significantly larger than the sun, rushed past (from the south); then such a cannonade erupted that all the workers in the boat rushed to hide in the cabin, forgetting the danger threatening from the rapids. The first blows were weaker, but then they grew louder. He estimated the sound effect lasted for about three to five minutes. The intensity of the sounds was so great that the boatmen were completely demoralized, and it took a great deal of effort to get them back into their positions in the boats.""

It looks from the account that the bluish light preceded the fiery body's passage. However, since this is a second-hand account, it's difficult to say for sure.

In any case, it can be said that at least one or two bodies were observed flying west of Kezhma. There is also evidence of similar activity east of Kezhma (please, note that at the time of the event the sun was almost strictly on the east).

The account (by Tikhon Naumovich Naumenko - see more info in [Ol’khovatov, 2023a]) was written in Moscow on Jan.21, 1936. Here are several fragments translated by the Author:

"The day was extremely sunny and so clear that we did not notice any cloudlet on the horizon; wind did not move, - the total silence.<...> ... at the moment when I got up from boards, among quickly amplifying sound of a thunder the first, rather weak blow sounded;<...> ... when I quickly turned in the direction of blow, beams of the sun were crossed (across) by a wide fiery white strip on the right side of beams, and with left towards the North (or if to take from Angara then - behind the Kezhemsky field) an irregular shaped, even more fiery white (paler that the sun, but almost identical with beams of the sun) little oblong body (in the form of a cloudlet, diameter is much more than the moon..., without the correct outlines of edges) flew to taiga.<...> After this second blow... the lump {‘cloudlet’ – A.O.} already disappeared, but the tail (to be more correct - a strip), already all came to be on the left side of beams of the sun, having cut them, and became many times wider, than was on the right side; and right there, through shorter period, than was between the first and second blow, the third thunderclap {blow – A.O.} and such strong (and as if with several blows merged together inside it), even with a crackling followed that all earth began to tremble, and such echo spread over taiga, and even not an echo, but some deafening continuous rumble was carried; it seemed that this rumble captured all taiga of immense Siberia. <...>

When we came to the settlement, we saw on streets the whole crowds of people ( both locals, and our companions exiled ) who were hotly discussing and on various ways interpreting this unusual phenomenon; because all our companions at the time of flight of a meteorite were in rooms, and some even slept, ... <...>

Our comrades, in their explanations, constructed a scenario involving the fall of a rare and unusually large meteorite to earth; they achieved this size thanks to the extraordinary force of the thunderclaps. Typically, especially when observing such flights in the evening or at night, we see only the fiery glint of the head and a rather long, comparatively narrow tail. In this case, however, the tail was wider than usual, and, thanks to the width, it seemed significantly smaller than we had seen at night. Perhaps it was the brightest sunlight of that day and the moment of the flight itself, which shortened the meteorite's tail."

Here is a sketch by Naumenko of the ‘flying cloudlet’ on Fig.3.

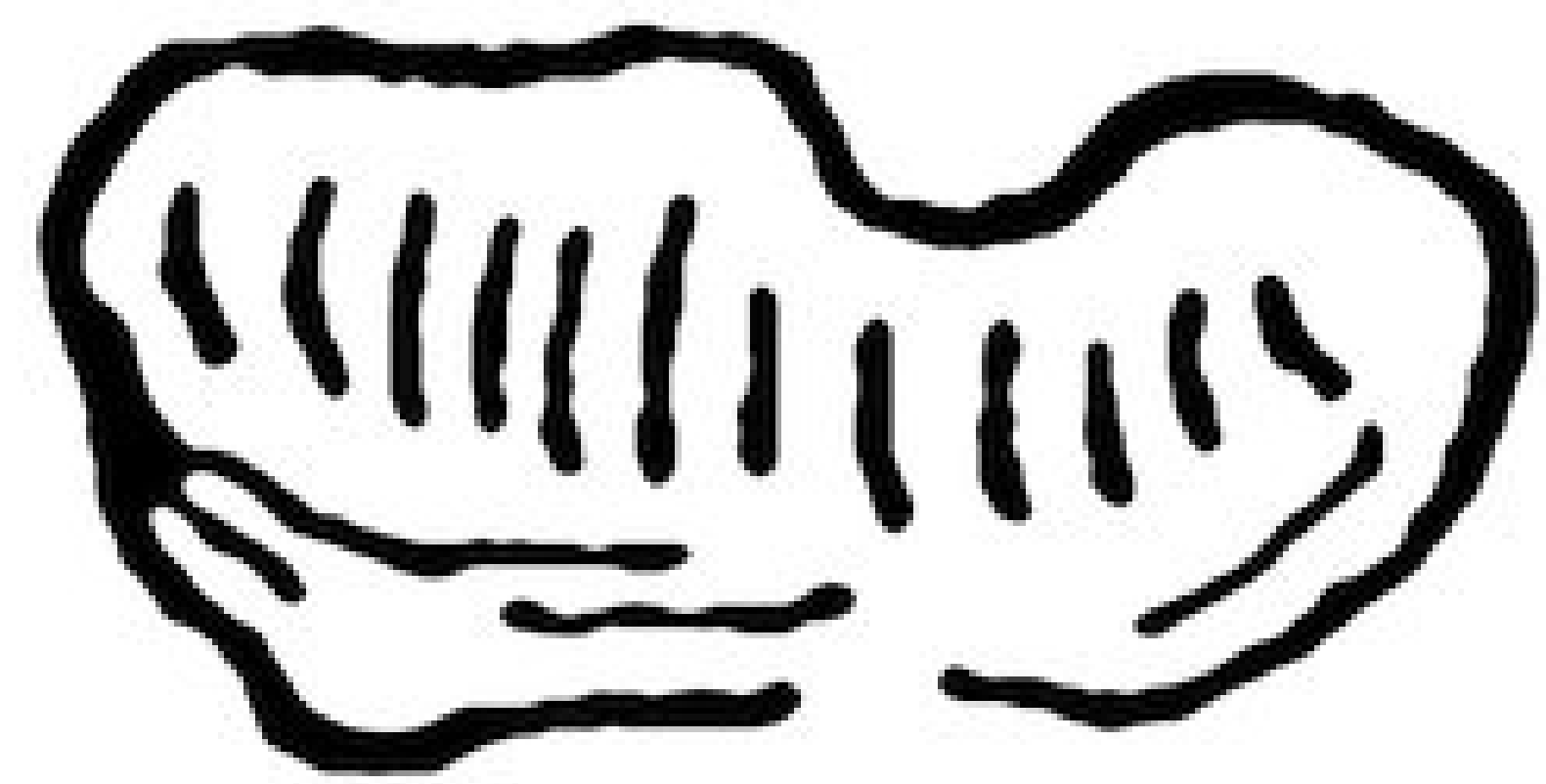

**Fig.3**

Thus, the flight of a luminous formation over Kezhma was immediately interpreted by the educated part of the population (mainly political exiles, who, by the way, even did not see the event) as a phenomenon known to them - a "meteorite fall", which was later formalized in the police officer's report. This is how the "Tunguska meteorite" was born (there was also a newspaper note about a fall of a meteorite near the town of Kansk).

By the way, the Naumenko's account also points that there was no cloud cover over Kezhma (Naumenko was located about half a kilometer from Kezhma) at the time of the Tunguska explosion. Cloud cover then began to rapidly develop. It's possible that the account by Naumenko reflects this phenomenon in its early stages.

And here is what happened to the east of Kezhma. The village of Mozgovaya was located approximately 10-12 km east of Kezhma. A response to a questionnaire distributed by Kulik in 1928 recorded an eyewitness account of a sighting in Mozgovaya [Vasil'ev, et al., 1981]. The account was written using old Russian slang, so here is paraphrased retelling of some main points in English (by A.O.) below.

An eyewitness was harrowing in an island with his helpers. They lay down to rest. Just as they fell asleep, a distant rumble began: "gro-gro-gro!...". "Something" in the sky moved northward: thump, thump, thump, and then it burst open and a strip of light appeared, and then again thump, thump, gr-r-r-ah! and again the light ripened. The horses fell to their knees. The eyewitness remembered the phenomenon so vividly that it was "as if it had happened yesterday".

It is noteworthy that a similar phenomenon, “pulsating glow in sync with powerful sounds,” was also observed in Kezhma. Here is from [Vasil'ev, et al., 1981], translated by A.O.:

"Bryukhanov A.K. was interviewed in 1929 by the teacher Z. Vostrikova, who gave her notes to L.A. Kulik. Lived in Kezhma.

" ... I haven't had time to get dressed yet after the bath, as I hear a noise. I jumped out into the street as I was and immediately threw a glance at the sky, because I could hear the noise from there. And I see: blue, green, red, hot (orange) stripes are going across the sky, they are as wide as the street.

The stripes went out and the rumble was heard again and the ground shook. Then again and again the stripes appeared and went "under the north". It seemed that they were about 20 versts from Kezhma. Well, then I heard that the end of them was far away, in the Tungus campsite.<…>"”

Colored stripes were also seen in Kezhma accompanying a flying red ball. In 1930 Ye.L. Krinov interviewed K.A. Kokorin. Here it is ([Vasil'ev, et al., 1981], translated by A.O.):

"Kokorin does not remember the exact day and year of the fall, but he remembers that three or four days before Petrov's day, at 8-9 o'clock in the morning, no later. The sky was completely clear, there were no clouds. He entered the bathhouse (in the courtyard), managed to take off his top shirt, when suddenly he heard sounds like cannon shots. He immediately ran out into the courtyard, open to the southwest and west. At this time, the sounds were still continuing, and he saw in the southwest, at an altitude of about half the distance between the zenith and the horizon, a flying red ball, and on the sides and behind his rainbow stripes were visible. The ball flew for 3-4 seconds, disappeared in the northeast (the directions were restored from memory when giving a reading of 1/I-1930 by compass). The sounds were heard during the flight of the ball, but they immediately stopped when the ball disappeared behind the forest.<…>"

It is remarkable that the ball does not seem to be bright and it did not leave any long-living trail (unlike the 2013 Chelyabinsk meteoritic event). Also it is important that K.A. Kokorin saw the ball in the southwest initially, i.e. in the other side of the world than the "cloudlet" that Naumenko saw.

Thus, eyewitnesses from Kezhma and its surrounding areas observed a wide variety of atmospheric phenomena. These included the passage of luminous formations both west and east of Kezhma. It is important to add that a similar variety of phenomena in the atmosphere was present in a number of other places - see, for example, [Ol’khovatov, 2023a; 2023b].

These atmospheric phenomena were observed during worsening weather, which was expressed in the appearance of the windstorm near Kezhma and the sharp increase in cloudiness.

It is important to note that on June 30, 1908 increased cloud cover occurred over a significant area to the north and northwest of Lake Baikal [Ol'khovatov, 2003].

On July 1, 1908 the Kezhma weather station recorded a weak rain and sounds of a distant thunderstorm. On July 2 the station recorded a much stronger rain, hail, and a thunderstorm in Kezhma, as well as sounds of a distant thunderstorm.

It is interesting that studies have shown [Gorbatenko, 2003], in June and July 1908 in Kezhma, a significant increase in the amount of precipitation was observed compared to the long-term norm.

## 3. Discussion

Initially, astronomical literature suggested that the alleged Tunguska spacebody passed over the Kezhma region, based on initial observations, which are used to be the most reliable. That is, the trajectory ran roughly from south to north. Then, in 1949, Krinov proposed a trajectory passing from the southeast. In 1955, astronomer N. N. Sytinskaya analyzed both trajectories and was unable to favor either, as both had sufficiently reliable evidence. Finally, in the 1960s, residents were interviewed on the banks of the Nizhnyaya Tunguska River who testified that the trajectory ran almost from east to west (see Fig.1, by the way, linking this trajectory to the Tunguska event requires a lot of assumptions and suppositions - see, for example, [Bronshten, 1999]). As this trajectory roughly corresponded to the direction of the axis of symmetry of the forestfall, this trajectory has since been generally considered in the astronomical literature as the most probable.

However, many researchers point out that this trajectory contradicts the very reliable accounts of eyewitnesses from Kezhma and its environs. Academician (of USSR Academy of Medical Sciences) Nikolai Vasil'ev, who was informal leader of the Tunguska research wrote [Vasil'ev, 1992] (a mistype is corrected, and please note that a reference [13] in the text is [Demin et al., 1984]):

> "Analysis of the catalog of statements by eyewitnesses to the disaster [11], the total number of which runs to a few hundred, reveals a fact that has not been clarified to date, namely that thunderlike sounds were heard not only during and after the flight of the bolide, but even before it. <...> It would hardly be realistic to explain them away as subjective errors, since claims of this kind are made over and over and independently of each other.
> <...>
>
> The second factor, a fairly odd factor, is related to the direction of motion of the body. Analysis of statements by witnesses who gathered along the hot tracks of the event [11] and in the 1920s and 1930s [25, 28] led the first investigators of the problem (L. A. Kulik, I. S. Astapovich, and E. L. Krinov) to the unanimous conclusion that the bolide traveled in the direction from south to north. However, analysis of the vector structure of the timber fall due to the shock wave of the Tunguska meteorite gives an azimuth of 114° [29, 30], and the field of burn damage even gives an azimuth of 95° [6-8], i.e., it indicates that the meteorite traveled from nearly east to west. It should be added that this direction also is confirmed by an analysis of the statements of eyewitnesses who lived at the time of the event in the upper reaches of the Lower Tunguska River (in the region of Preobrazhenka, Erbogachen, and Nepa).

The inconsistency is obvious. There have been repeated attempts to explain it from various standpoints. In particular, the hypothesis has been advanced [13] that on June 30, 1908 not one but several bolides flew over Central Siberia. However, this interpretation appears to be quite a stretch because among the many hundreds of documented statements by eyewitnesses, there is not a single one in which two bolides observed on the same day are mentioned, although an overlap of zones of visibility is more than likely in the case in question. The version espoused by F. Yu. Zigel' concerning a maneuver by the TM in earth's atmosphere raised a great controversy. However, it may be discussed seriously only if one assumes that the Tunguska cosmic body was technogenic in origin."

Vasil'ev also discussed this with more details in 1994 - see, for example, in [Ol'khovatov, 2023a].

In this regard, the Author would like to present a conversation that occurred on the sidelines of an international conference on Tunguska in Krasnoyarsk in 1998. The conversation between Vasil'ev and the Author went something like this.

- Vasil'ev: So you think there were several fireballs?
- The Author: There were many.
- Vasil'ev: But none of the eyewitnesses saw two fireballs!
- The Author: The altitudes are different there.
- Vasil'ev: A-a-a-h!...{meaning "understand"  -A.O.}

The eyewitness accounts are in agreement with the geophysical interpretation of the Tunguska event, according to which the luminous formations could be at very low altitude - even near the ground. For example, Kezhma was a rather large village (1248 residents on Jan.1, 1911) stretching along the northern bank of the river in appr. E-W direction. So a low-flying luminous formation seen at one part of Kezhma could be missed at another part.

Other similar reports (gotten right after the Tunguska event) of fireballs seen just in localized areas and flying in various directions can be found, for example, in [Ol'khovatov, 2023a].

In addition to the diversity of optical phenomena, the geophysical interpretation can also explain a number of other peculiarities of the Tunguska event. however this is beyond the scope of this paper.

In any case, the data from the Kezhma meteorological station are consistent with eyewitness accounts of rain and thunderstorm after the Tunguska event in the vicinity of the epicenter, and also with the words of residents of the Angara region that after the fireball fell into the taiga, a very powerful, unprecedented thunderstorm with heavy rain soon broke out - see [Ol'khovatov, 2023b] for details.

Remarkably, that some researchers of the tree-damage in Tunguska also admit the rain and even lightnings [Ol'khovatov, 2023b; 2025f].

In 2008 a meteorologist from Tomsk (who took part in the KSE expeditions) presented results of a new meteorological analysis [Gorbatenko, 2008]. Here it is [Gorbatenko, 2008] (TKT is the alleged Tunguska space body, translated by A.O.):

> "Judging by the changes in air temperature and pressure at stations located around the place where the TKT fell and during the fall, it can be assumed that over the territory of the south-west of the Eastern (and partly in the southeast of Western) Siberia there was an extensive cyclone. Its warm front sequentially passed through the settlements: Kansk, Yeniseisk, Kezhma. The center of the cyclone was located east of Krasnoyarsk, in the Kansk region, with a pressure in the center of about 735 mm Hg (980 mb).
>
> In addition, analyzing the air temperature values (Fig. 2) at stations close to the area of interest to us, it is safe to state that in the period from 06/30/1908 to 07/01/1908 above the territory of interest to us passed a warm front and for 0.5-1 days the territory was in a warm sector of the cyclone, and therefore at this time there could be stratus clouds and a general deterioration visibility. A little later it was replaced by a cold front with cumulus clouds and thunderstorms. Therefore, when analyzing other effects observed in the atmosphere during these days, it is necessary to take into account all the weather variations observed during the day when the TKT fell."

In the opinion of the Author, the data of the Kezhma weather station and the analysis by Gorbatenko, as well as the eyewitness accounts, are in agreement with the geophysical interpretation of the 1908 Tunguska event.

In any case, the meteorological factor is an important aspect of the Tunguska event, and so it deserves detailed research.

## 4. Conclusion

The note by the observer of the weather station in Kezhma, concerning, in particular, the appearance of the two huge fiery circles is very important. It is an addition to the diversity of optical phenomena reported from Kezhma and its environs.

The weather log of the weather station shows that the Tunguska event occurred during the sharp increase in cloud cover and deteriorating weather. At the beginning of July, rains and thunderstorms began, including hail, according to the weather log. The data of the weather station is consistent with eyewitness accounts and results of the meteorological analysis published in 2008 by the meteorologist from Tomsk - V.P. Gorbatenko.

In the opinion of the Author, the data of the Kezhma weather station and the analysis by Gorbatenko, as well as the eyewitness accounts, are in agreement with the geophysical interpretation of the 1908 Tunguska event.

In any case, the meteorological factor is an important aspect of the Tunguska event, and so it deserves detailed research.

The general conclusion is that the Tunguska event was a very complex phenomenon. Research of the Tunguska event requires the participation of experts in various fields. In the opinion of the Author, researching the Tunguska event will allow us to better understand the life of such a complex system as our planet.

**ACKNOWLEDGEMENTS**

The Author wants to thank the many people who helped him to work on this paper, and special gratitude to his mother - Ol'khovatova Olga Leonidovna (unfortunately she didn't live long enough to see this paper published...), without her moral and other diverse support this paper would hardly have been written.

## References

Astapowitsch, I.S. (1940). New data concerning the fall of the great [Tungus] Meteorite on June 30, 1908, in Central Siberia. // Popular Astronomy, Vol. 48, p.433. https://doi.org/10.1111/j.1945-5100.1940.tb00319.x

Bronshten, V. A. (1999). Trajectory and orbit of the Tunguska meteorite revisited. // Meteoritics & Planetary Science, vol. 34, p. A137-A143.

Demin, D.V., Dmitriev, A.N., Zhuravlev, V.K. (1984). Informatsyonnyi aspect issledovanii Tungusskogo fenomena 1908 g. // Meteoritnye issledovaniya v Sibiri. Novosibirsk, Nauka, pp.30 - 49 (in Russian).

Gorbatenko, V.P. (2003). O klimaticheskikh anomaliyakh iyulya 1908 goda.// Konferentsyya “95 let Tungusskoi probleme”. Moscow, MSU, 23-24 iyunya 2003 g., ch.1. (in Russian). https://tunguska.tsc.ru/ru/science/conf/2003/p1/gorbatenko/

Gorbatenko, V.P. (2008). Ob anomalii osadkov, vypavshykh v 1908 gody v Krasnoyarskom krae. //Fenomen Tunguski: mnogoaspektnost problemy. Novosibirsk, OOO IPF "Agros", pp.142-144. ISBN 978-5-9657-0130-8 (in Russian).

Ol'khovatov, A.Y. (2003). Geophysical Circumstances Of The 1908 Tunguska Event In Siberia, Russia. // Earth, Moon, and Planets 93, 163–173. https://doi.org/10.1023/B:MOON.0000047474.85788.01

Ol'khovatov, A. (2020a). New data on accounts of the 1908 Tunguska event.// Terra Nova,v.32, N3, p.234.
https://doi.org/10.1111/ter.12453

Ol'khovatov, A. (2020b). Some comments on events associated with falling terrestrial rocks and iron from the sky.// https://arxiv.org/abs/2012.00686
https://doi.org/10.48550/arXiv.2012.00686

Ol'khovatov, A. (2021) - The 1908 Tunguska event and forestfalls. // arXiv:2110.15193 ; https://doi.org/10.48550/arXiv.2110.15193

Ol'khovatov, A. (2022) - The 1908 Tunguska Event And The 2013 Chelyabinsk Meteoritic Event: Comparison Of Reported Seismic Phenomena. // eprint arXiv:2206.13930 , https://arxiv.org/abs/2206.13930 ;
https://doi.org/10.48550/arXiv.2206.13930

Ol’khovatov, A. (2023a) - The 1908 Tunguska event: analysis of eyewitness accounts of luminous phenomena collected in 1908. // arXiv:2310.14917 ;
https://doi.org/10.48550/arXiv.2310.14917

Ol’khovatov, A. (2023b) - The Evenki accounts of the 1908 Tunguska event collected in 1920s – 1930s. // arXiv:2402.10900 ;
https://doi.org/10.48550/arXiv.2402.10900

Ol’khovatov, A. (2025a). The 1908 Tunguska Event and Bright Nights. // arXiv:2502.01645, https://arxiv.org/abs/2502.01645 ;
https://doi.org/10.48550/arXiv.2502.01645

Ol’khovatov, A. (2025b). Some mechanical and thermal manifestations of the 1908

Tunguska event near its epicenter. // https://eartharxiv.org/repository/view/8790/ ; https://doi.org/10.31223/X52F0H

Ol'khovatov, A. (2025c). Some Historical Misconceptions and Inaccuracies Regarding The 1908 Tunguska Event. // https://arxiv.org/abs/2505.05484 ; https://doi.org/10.48550/arXiv.2505.05484

Ol'khovatov, A. (2025d). The 1993 Jerzmanowice event in Poland and the 1908 Tunguska event. // https://eartharxiv.org/repository/view/9435/ ; https://doi.org/10.31223/X59F0P

Ol'khovatov, A. (2025e). Some Geophysical Aspects Of The 1908 Tunguska Event. // http://arxiv.org/abs/2507.21296 ; https://doi.org/10.48550/arXiv.2507.21296

Ol'khovatov, A. (2025f). The 1908 Tunguska event and electromagnetic phenomena. https://eartharxiv.org/repository/view/9781/ ; https://doi.org/10.31223/X5RR0T

Ol'khovatov, A. (2025g). The 1908 Tunguska event and the atmospheric optical anomalies. // https://arxiv.org/abs/2510.03239 ; https://doi.org/10.48550/arXiv.2510.03239

Ol'khovatov, A. (2025h). On radiocarbon near the epicenter of the 1908 Tunguska event. // https://eartharxiv.org/repository/view/10610/ https://doi.org/10.31223/X5V46T

Ol'khovatov, A. (2025i). The 1908 Tunguska event and some distant phenomena. // https://arxiv.org/abs/2511.17549 ; https://doi.org/10.48550/arXiv.2511.17549

Ol'khovatov, A. (2026a). The 1935 Guyana event and the 1908 Tunguska event. https://eartharxiv.org/repository/view/11337/ https://doi.org/10.31223/X5645J

Ol'khovatov, A. (2026b). The 1908 Tunguska event and some mini-Tunguskas. https://eartharxiv.org/repository/view/12177/ https://doi.org/10.31223/x5h473

Rykatchew, M. (1911). Annales de l'observatoire physique central Nicolas, Annee 1908, Peterburg. (in French).

Vasil'ev, N.V. (1992). Paradoxes of the problem of the Tunguska meteorite. // Soviet Physics Journal 35, pp. 294–299. https://doi.org/10.1007/BF00895778

Vasil'ev N.V., Kovaleskii A.F., Razin S.A., Epiktetova L.E.(1981). Pokazaniya ochevidtsev Tungusskogo padenia. - Tomsk, TGU,  N5350 - 81Dep, 304 p. (in Russian).